\documentclass[fleqn,usenatbib]{mnras}

\usepackage{graphicx}	
\usepackage{amsmath}	
\usepackage{amssymb}	

\usepackage{newtxtext,newtxmath}

\usepackage[T1]{fontenc}
\usepackage{ae,aecompl}

\usepackage{float}
\restylefloat{table}
\usepackage{subcaption}

\usepackage[dvipsnames,svgnames,x11names]{xcolor}
\usepackage[final]{changes}
\makeatletter
\@namedef{Changes@AuthorColor}{BrickRed}
\colorlet{Changes@Color}{BrickRed}
\makeatother

\title[Accuracy vs. Complexity]{ Accuracy vs. Complexity: Calibrating radio interferometer arrays with non-homogeneous element patterns }

\author[Jones and Wayth]{
Jake L. Jones,$^{1}$\thanks{E-mail: jake.jones@curtin.edu.au}
Randall B. Wayth,$^{1}$
\\
$^{1}$ICRAR - Curtin University. GPO Box U1987, Perth, 6845. Australia\\
}

\date{Accepted XXX. Received YYY; in original form ZZZ}

\pubyear{2021}

\begin{document}
\label{firstpage}
\pagerange{\pageref{firstpage}--\pageref{lastpage}}
\maketitle

\begin{abstract}
Radio interferometer arrays with non-homogeneous element patterns are more difficult to calibrate compared to the more common homogeneous array. In particular, the non-homogeneity of the patterns \added{has} significant implications on the computational tractability of evaluating the calibration solutions.
We apply the A-stacking technique to this problem and explore the trade-off to be made between the calibration accuracy and computational complexity.
Through simulations\added{,} we show that this technique can be favourably applied in the context of an SKA-Low station\added{.}
\added{We show that the minimum accuracy requirements can be met at a significantly reduced computational cost, and this cost can be reduced even further if the station calibration timescale is relaxed from 10 minutes to several hours.}
We demonstrate the impact antenna designs with differing levels of non-homogeneity have on the overall computational complexity in addition to some cases where calibration performs poorly.

\end{abstract}


\section{Introduction}

The past decade has seen a proliferation of new and upgraded radio telescopes that operate as phased-arrays, especially at low radio frequencies (approximately 20-300\, MHz).
These radio telescopes typically have very large fractional bandwidths, which present some unique challenges for phased-arrays.
To avoid potential problems with grating lobes and \deleted{scan blindness} \added{nulls within the beam pattern} (caused by mutual electromagnetic coupling between array elements) associated with regularly-spaced antenna arrays, some telescope designers have opted for a pseudo-random configuration of the antenna elements in the array, e.g. the low band of LOFAR \citep{2013A&A...556A...2V}, and the LWA \citep{2009IEEEP..97.1421E}.
The future Square Kilometre Array\added{,} Low Frequency Array (SKA-Low) will have phased-array ``stations'' that also have pseudo-randomly configured antenna elements for the reasons noted above. First generation prototype stations for SKA-Low, including the Engineering Development Array \citep[`EDA'][]{2017PASA...34...34W} and \added{Aperture Array Verification System 1} ('AAVS1' Benthem et al., in prep.), used the same pseudo-random configuration.

A key science driver of the new generation of low frequency radio telescopes is centred around detecting faint signals from neutral Hydrogen in the early universe \citep{2006PhR...433..181F} -- the Cosmic Dawn (CD) and Epoch of Reionisation (EoR).
The expected signals are 4 to 5 orders of magnitude fainter than ``foreground'' Galactic and extragalactic radio sources, hence the telescopes are required to achieve very high \deleted{fidelity} \added{dynamic range} in calibration and associated data processing (generating images and/or power spectra) so that artifacts from stronger foreground sources do not contaminate the fainter signals of interest.
The observable quantity for CD/EoR science is also fundamentally a spectral line signal, \added{therefore the} accuracy of bandpass calibration and source subtraction as a function of frequency is critically important \citep[see ][ for a recent review]{2019cosm.book.....M}.

The issues underlying calibration/imaging requirements for early universe science are fundamentally the same issues for what has traditionally been called ``high dynamic range'' imaging in radio interferometry. I.e., calibrating and modelling \added{the} data with sufficient accuracy that faint sources are not obscured by artefacts caused by bright sources.
Equivalently, it can be viewed as processing the data with sufficient accuracy that long integrations of data remain limited by random (thermal noise) errors.
Achieving \added{a} high dynamic range implies reduced systematic error, which is intimately related to the quality of antenna models and \added{the} sky model used in calibration and imaging.

For example, artefacts caused by az/el mount dish-based antennas have been known for some time to cause artefacts in radio telescope imaging software because the antenna feed legs cause asymmetry in the beam pattern\added{. T}hus breaking a key simplifying assumption traditionally used in radio astronomy: that the antenna beam patterns are all identical and constant on the sky during an observation.
Similarly, the wide field-of-view of some instruments causes baseline-dependent phase screens (``w-terms'') to break the assumption of identical antennas.
The need to incorporate changing antenna beam patterns into calibration/imaging/deconvolution of radio astronomy data has led to new algorithms and software tools \citep[e.g.][]{2008A&A...487..419B,2012ApJ...759...17S,2013ApJ...770...91B,2014MNRAS.444..606O,A-stack,2018A&A...616A..27V,2019ApJ...874..174P,2015MNRAS.449.2668S,2018A&A...611A..87T}.

In the context of low frequency phased-arrays, including SKA-Low, the coherently combined station beam from the fixed (on the ground) pseudo-random antenna array represents the worst case of ``non-identical, non-constant'' antenna beam patterns. The calibration task for SKA-Low has been split into two distinct parts; the first is standalone calibration of each station, which forms one or more station beams; the second is array-level calibration, which treats each station beam like a single large antenna.

The first of these calibration tasks -- standalone calibration of a station -- is the focus of this paper. The previous and current generation of ``SKALA'' log-periodic dipole antennas in the baseline design for SKA-Low \citep{SKALA_description,SKALA_characterisation} have been shown to be strongly mutually electromagnetically coupled \citep{SKALA_mutual_coupling}. In order to form a coherent station beam, the station must be calibrated, which can be achieved with astronomical radio sources using standard calibration techniques (which incorporate the mutual coupling) by using the station as a standalone interferometer, rather than a phased-array.
The calibration process requires model visibilities, which are generated using a model of the sky and a model of the antenna radiation patterns.
Importantly, the strong mutual coupling between antennas \added{cause} each antenna in the array to have a unique radiation pattern, or beam, on the sky.

\added{
Calibration techniques that incorporate non-homogeneous element patterns can exhibit a significant increase in computational complexity in addition to requiring accurate knowledge of each element pattern. Therefore, non-homogeneous element patterns can have significant computational, hence cost, impact. The increase in computational cost could easily be overlooked in the design and costing process of a radio telescope, both for the up-front modelling and operations perspective, as this cost is usually negligible.
}

Current thinking for SKA-Low is for all stations to be uniquely randomised. Taken at face value, this implies that knowledge of the radiation pattern for each antenna element, in each station, for all frequencies of interest, over the entire sky, is required.
It may yet prove to be computationally intractable to simulate \added{the antenna patterns of} an SKA-Low comprised of truly uniquely random station layouts, however\added{,} even if it is possible, there remains the related issue of how to use the simulated antenna beams in the calibration process.
That is, the evaluation of model visibilities based on the simulated heterogeneous antenna beams is itself an intractable problem when applied on a large scale such as the SKA-Low.
\added{In this paper, we explore a novel technique that can reduce the computational complexity of this problem while maintaining the target accuracy requirements, resulting in a more tractable problem.
We explore this technique in the context of real-time calibration for the SKA-Low prototype stations AAVS1 and EDA2, although the methods used can be applied to any radio interferometer.}

\section{Self Calibration with Non-Homogeneous Element Patterns}

In general, the calibration problem for an interferometer reduces to solving the following minimisation problem for the complex-valued direction independent gains\added{,}
\begin{equation}
  \mathbf{g} = \underset{ \mathbf{g} }{ \text{argmin} } \ \Big{|}\Big{|} \mathbf{\tilde{R}} - \mathbf{G} \mathbf{R} \mathbf{G}^H \Big{|}\Big{|}_F^2 \added{,} \label{eqn:Cal}\\
\end{equation}
\noindent where \added{$||.||_F$ denotes the Frobenius norm,} $\mathbf{\tilde{R}}$ is the measured \added{coherence matrix}, $\mathbf{R}$ is the model \added{coherence } matrix and $\mathbf{G}$ is a diagonal matrix containing the direction independent antenna gains $\mathbf{g}$. \deleted{Additionally, }\added{T}he auto-correlations along the diagonals of $\mathbf{\tilde{R}}$ and $\mathbf{R}$ are typically set to zero because of the increased level of self-correlated noise.
A number of non-linear optimisation methods can be employed to solve such a problem, of which the Levenberg-Marquardt algorithm is one of the most commonly used in radio astronomy and therefore used throughout this paper.

The self-calibration procedure depends on \emph{a-priori} knowledge of the model \added{coherence} matrix $\mathbf{R}$. Unfortunately the estimation of $\mathbf{R}$ is non-trivial, especially for arrays with large fields of view and non-homogeneous element patterns \citep{CAL-DDE}.
Since the self-calibration algorithm assumes $\mathbf{R}$ is error free it follows that any error in its estimation will lead to the introduction of a systematic bias within the calibration solution. 
\added{For a noiseless and well-conditioned} system, one would expect the calibration solution to have magnitude gains equal to unity and have zero phase.
Any deviation from the expected solution is a measure of the systematic \added{error} that results from an imperfect model \added{coherence} matrix.

The model \added{coherence} matrix $\mathbf{R}$ is evaluated using the Radio Interferometer Measurement Equation (RIME) \citep{RIME} which describes the relationship between the sky brightness distribution $B(l,m)$ and the visibilities $V_k$ as follows\added{,}
\begin{align}
   V_{k} = \iint A_{k}(l,m) B(l,m) \ e^{-i 2 \pi (u_{k} l + v_{k} m + w_{k} (n-1) )} \ \frac{ dl dm }{ n } \added{,} \label{eqn:RIME}
\end{align}

\noindent where $ n = \sqrt{ 1 - l^2 - m^2 } $ \added{and} $k$ denotes the baseline index\added{. The term} $A_{k}(l,m)$ is the \added{complex-valued} baseline dependent beam pattern and $l$, $m$ \& $n$ are direction cosines corresponding to coordinates $u$, $v$ \& $w$ respectively. \added{For simplicity,} notation for frequency dependence has been dropped in addition to the antenna polarisation (X or Y pol). It is assumed that the sky is un-polarised and therefore the cross polarisation terms can be ignored. 
\added{The calibration problem is well-conditioned if both $A_k(l,m)$ and $B(l,m)$ are non-zero for all $k$, in which case the calibration solution of a noiseless system would be unity.}

The simplest approach to calculate the model \added{coherence} matrix is to directly evaluate the RIME through the use of the \added{Discrete} Fourier Transform (DFT) method. This involves computing the 2D DFT with an image size of $N_P \times N_P$ pixels for every single baseline, which for a 256 element interferometer results in a huge number of $K=32640$ baselines. \added{This method yields the exact model coherence matrix and has complexity,}
\begin{align}
    C_{\text{DFT}} &=  \mathcal{O}\Big( \ N_p^2 \ K \ \Big) \added{.}
    \label{eqn:C_DFT}
\end{align}
\added{Unfortunately, this complexity leads to the methods intractability in practice.}

\section{Efficient Evaluation of the RIME}

\added{Conventional radio interferometers with a small field of view and steerable elements can significantly simplify the RIME because the sky brightness distribution can be considered as a point source. Therefore, for each baseline, only a single expression needs to be evaluated as opposed to a full 2D integration over the sky. 
Modern instruments such as the SKA-Low use many phased-arrays (stations) that must individually undergo calibration as standalone interferometers. Since a station consists of a fixed array of antennas, each with a large field of view that always points at the zenith, the full diffuse structure of the sky must be taken into account. Therefore, the RIME cannot be simplified in the typical manner.}

For an interferometer array with homogeneous element patterns the term $A_k(l,m)$ which is generally a baseline dependent term\added{,} no longer depends on the baseline index, reducing the RIME to a form that can exploit the computational efficiency of the Fast Fourier Transform (FFT). \added{However,} the FFT computes visibilities on a regular grid and since the baseline locations in \added{the} (u,v) domain may not fall exactly on a grid point\added{,} there is an additional interpolation step that is required to obtain the proper result. This step usually takes the form of a convolution between a small de-gridding kernel and the output of the FFT. Because the overhead cost of this de-gridding step is negligible compared to the benefit of using the FFT, the overall complexity is significantly reduced compared to the DFT method. Historically, radio interferometers have exploited this method to great advantage\added{,} however modern instruments such as the SKA-Low are pushing the boundaries \added{by using antennas with} non-homogeneous element patterns and must find another solution.

A number of algorithms have been been devised that aim to reduce the complexity of evaluating the RIME, these include but are not limited to the Average Embedded Element \citep['AEE'][]{AEEvFEE}, AW-projection \citep{AW-proj} and A-stacking methods \citep{A-stack}. A common feature of these algorithms is the trade-off between their accuracy and the computational complexity. 
\added{Therefore, in order to reduce the computational complexity, a certain level of systematic error must be tolerated. }

The simplest of these algorithms is the Average Embedded Element method (AEE)\added{,} which \deleted{initially} takes the average of all the Embedded Element Patterns (EEP's) using \added{the equation,}
\begin{align}
    \Bar{A} = \Big( \frac{1}{N} \sum_{i=0}^{N} ||\mathbf{E}_i|| \Big)^2 \added{\text{,}}
    \label{eqn:AEE}
\end{align}
\added{where $\mathbf{E}_i$ is the complex-valued embedded element pattern for the $i$'th antenna out of $N$ total antennas.}
The average pattern is then applied for all elements in the array, therefore removing any baseline dependence from the integral. As was the case for arrays with homogeneous element patterns, this method allows for the use of the FFT in combination with a de-gridding step. \added{For non-coplanar arrays, W-projection \citep{W-proj} can account for the extra w-term in the form of a convolution kernel often combined with the de-gridding kernel.}
\added{The AEE method leads to a significant improvement in computational complexity given by,}
\begin{align}
    C_{\text{AEE}} &= \mathcal{O}\Big( \ N_p^2 + N_p^2 \ \log_2(N_p) + K N_{gW}^2 \ \Big)  \added{\text{,}}
    \label{eqn:C_AEE}
\end{align}
\added{where $N_{gW}$ is the size of the de-gridding kernel and W-projection kernel combined.}
Unfortunately, while the AEE method is the least costly algorithm\added{,} it is not particularly effective for arrays with vastly non-homogeneous element patterns as demonstrated by \cite{AEEvFEE}.
Due to the prevalence of the AEE method, its computational costs are well understood and therefore makes a good reference for comparison when evaluating more advanced methods.

AW-projection is an extension of W-projection that includes direction dependent beam effects \citep{AW-proj}. This method exploits the convolution theorem by shifting the baseline dependent corrections into the uv domain resulting in the computational complexity\added{,}
\begin{align}
    C_{\text{A-proj}} &=  \mathcal{O}\Big( \ N_p^2 \ \log_2 N_p + K N_{gA}^2  \ \Big)  \added{,}
    \label{eqn:C_Aproj}
\end{align}
\added{where $N_{gA}$ is the size of the convolution kernel.} A reduction in complexity is achieved if the baseline dependent term $A_k(l,m)$ has limited support in the uv domain, that is\added{,} most of the power is contained within a small area such that the size of the convolution kernel can be truncated to size $N_{gA}$ without introducing error. While this algorithm has proven to be useful for calibrating LOFAR \citep{A-proj-LOFAR} due to its smooth primary beam, arrays that exhibit beam patterns with significant spatial structure will derive limited computational benefit.

\subsection{A-stacking}
The A-stacking algorithm outlined in \cite{A-stack} is \added{a} very promising approach for efficiently evaluating the RIME\added{. T}he key idea behind the algorithm is to model the baseline dependent beam patterns as a linear combination of component functions \added{as follows,}
\begin{align}
    A_{k}(l,m) = \sum \limits_{i=0}^{N_B - 1} c_{i,k} \ f_i(l,m) \added{,}
    \label{eqn:Amodel}
\end{align}
\noindent where $f_i(l,m)$ and $c_{i,k}$ are the component functions and coefficients respectively and $N_B$ is the number of component functions used in the model. Typically\added{,} the greater the number of components used, the more accurate the representation becomes. Substituting this through the RIME we obtain \added{the equation,}
\begin{align}
    V_{k} = \sum \limits_{i=0}^{N_B - 1} c_{i,k} \iint f_i(l,m) \ B(l,m) \ e^{-i 2 \pi (u_{k} l + v_{k} m + w_k(n-1) )} \ dl dm \added{,} \label{eqn:A-stack} 
\end{align}
\added{where the $\frac{1}{n}$ term has been absorbed into the $A_k(l,m)$ approximation.}
\added{Since the main} integral no longer contains any baseline dependent terms\added{, we can now utilise} the FFT but at the cost of evaluating it $N_B$ times.
\added{Again, the w-terms are accounted for as a convolution during the de-gridding stage.} This leads to the computational complexity\added{,}
\begin{align}
    C_{\text{A-stack}} &= \mathcal{O}\Big( \  N_B( \ N_p^2 + N_p^2 \ \log_2(N_p) + K N_{gW}^2 \ )  \ \Big) \added{.}
     \label{eqn:C_Astack}
\end{align}
It should be noted that the integral term has the same complexity as the AEE method\added{,} leading to a proportional relationship where the complexity of A-stacking is $N_B$ times that of the AEE method. Since the parameter $N_B$ is linked to both the accuracy of the beam model and the computational complexity of the algorithm, this parameter can be tuned in order to meet \added{our} requirements.

\subsection{Modelling the Baseline Dependent Beam Patterns}

In principle, we want to develop an accurate model of the baseline dependent beam patterns using as few component functions as possible. The number of components required to meet the target level of accuracy will ultimately depend on the properties of the Embedded Element Patterns (EEP) in the array. 
Previous works have explored using low order models to represent individual EEP's using spherical waves, Zernike polynomials and the like \citep{6046372,6632431}. In \added{our} case\added{,} we wish to obtain a low order model of the baseline dependent patterns as opposed to \added{the} individual element patterns.
\added{
For convenience, we define the exact beam model as $\mathbf{A} = [ A_0 \ A_1 \ ... \ A_K ]$ and the approximate beam model as $\mathbf{M} = [ M_0 \ M_1 \ ... \ M_K ]$. Each column contains the unraveled pattern of the exact model $A_k(l,m)$ and the approximate model $M_k(l,m)$ respectively for a specific baseline, resulting in matrix dimensions $N_p^2 \times K$.}
A measure of the modelling error can therefore be expressed by $ || \mathbf{A} - \mathbf{M} ||_F $ where $F$ denotes the Frobenius norm. Under this formulation the linear combination of component functions as per equation \ref{eqn:Amodel} can be represented through a matrix multiplication of the form\added{,}
\begin{align}
    \mathbf{M} = \mathbf{F} \mathbf{C} = [ f_0 \ f_1 \ ... \ f_{N_B} ] [ c_0 \ c_1 \ ... \ c_K ] 
\end{align}
where the component functions $\mathbf{F}$ has dimensions of \added{$N_p^2 \times N_B$} and the coefficients $\mathbf{C}$ has dimensions \added{$N_B \times K$}. Note that due to this formulation, it is obvious that the beam model $\mathbf{M}$ has a rank that is equal to the number of components $N_B$.

The Singular Value Decomposition (SVD) of the matrix $\mathbf{A}$ is given by $\mathbf{A} = \mathbf{U} \mathbf{\Sigma} \mathbf{V^H}$\added{,} where $\mathbf{U}$, $\mathbf{\Sigma}$ and $\mathbf{V^H}$ are the left singular vectors, singular values and right singular vectors of the SVD respectively\added{.} The Eckart-Young-Mirsky theorem states that the best rank n approximation of $\mathbf{A}$ is given by \added{the} Truncated SVD\added{,}
\begin{align}
     \mathbf{M} = \mathbf{U_n} \mathbf{\Sigma_n} \mathbf{V^H_n} \added{,}
\end{align}
\added{where the subscript $n$ denotes the truncated form that includes only the largest $n$ singular values and corresponding vectors.}

Combining the left singular vectors and singular values together we arrive at our desired representation $\mathbf{M} = \mathbf{F} \mathbf{C}$ where the component functions are defined by $\mathbf{F} = \mathbf{U_n} \mathbf{\Sigma_n}$ and the coefficients are $\mathbf{C} = \mathbf{V^H_n}$. Given a fixed number of components, the Eckart-Young-Mirsky theorem guarantees that the model obtained via the Truncated SVD will minimise the modelling error $ || \mathbf{A} - \mathbf{M} ||_F $. Therefore\added{,} the Truncated SVD is used to generate the beam models used throughout this work.

\subsection{Application to the SKA-Low Station}

There are two antenna designs under consideration for the SKA-Low, the first is the SKA Log-periodic Antenna (SKALA) \citep{SKALA_description,SKALA_characterisation} and the second is based on the Murchison Widefield Array (MWA) dipole or "bow-tie" antenna \citep{MWA_description}. The SKALA antenna has gone through several design iterations, the design used in Aperture Array Verification System 1 (AAVS1) is the SKALA2, the Engineering Development Array 2 (EDA2) uses the bow-tie antenna\added{. Both the AAVS1 and EDA2} arrays have the same layout but differ by antenna design.

Evaluating the RIME requires \emph{a-priori} knowledge of the baseline dependent power patterns which in turn depend on the individual response of each antenna within the array. The Embedded Element Pattern (EEP) for each antenna has therefore been evaluated using electromagnetic simulation software called FEKO \citep{AAVS1.5_FEKO,FEKO-AAVS2}.
\added{There is an ongoing effort to verify these simulations empirically \citep{AAVS0.5_drone_verification}, however, within the scope of this paper it is assumed that these are error free.
Depending on the convergence rate of the solver, the time to simulate a 256 element array at a single frequency can range from days to weeks, even when using a relatively powerful server (Dual Xeon Platinum 8180 with 56 cores and 1.5TB RAM) as specified in} \cite{FEKO-AAVS2}.
\added{Accurately simulating the EEP's is therefore extremely computationally costly. Fortunately, this simulation need only be done once for each frequency and antenna layout, however, if the SKA-Low chooses to have a unique layout for each station, this task may become prohibitively expensive.
A conservative estimate of the compute time, assuming only 7 frequency points are required with 24 node hours of compute time per point, results in over 9 years of processing time for a single server (same specifications as above) to simulate 512 unique stations. This would require a significant amount of expensive server-grade equipment to achieve such a task in a reasonable amount of time.
However, once the EEP's have been simulated, they must be used within the calibration process which itself can be computationally intractable.
}

The baseline dependent beams $A_k(l,m)$ are a result of the element patterns that make up that baseline pair; assuming the sky is predominantly un-polarised these beams can be represented as an inner product of the two elements. FEKO outputs Jones matrices for each element in a radial polarisation basis ($\theta$, $\phi$)\added{.} \deleted{Therefore} The baseline dependent beams are calculated \added{using,}
\begin{align}
    A_{k} = \mathbf{E}_p \mathbf{E}^H_q = E_{\theta, p} E_{\theta, q}^* + E_{\phi, p} E_{\phi, q}^* \added{,}\label{eqn:EEP}
\end{align}
where \added{$\mathbf{E}_i$ is the Jones matrix representing the EEP of the i'th element,} 
$k$ is the baseline index and $p$ \& $q$ are the elements that make up that baseline pair.

\begin{figure*}
\centering
\includegraphics[width=\linewidth]{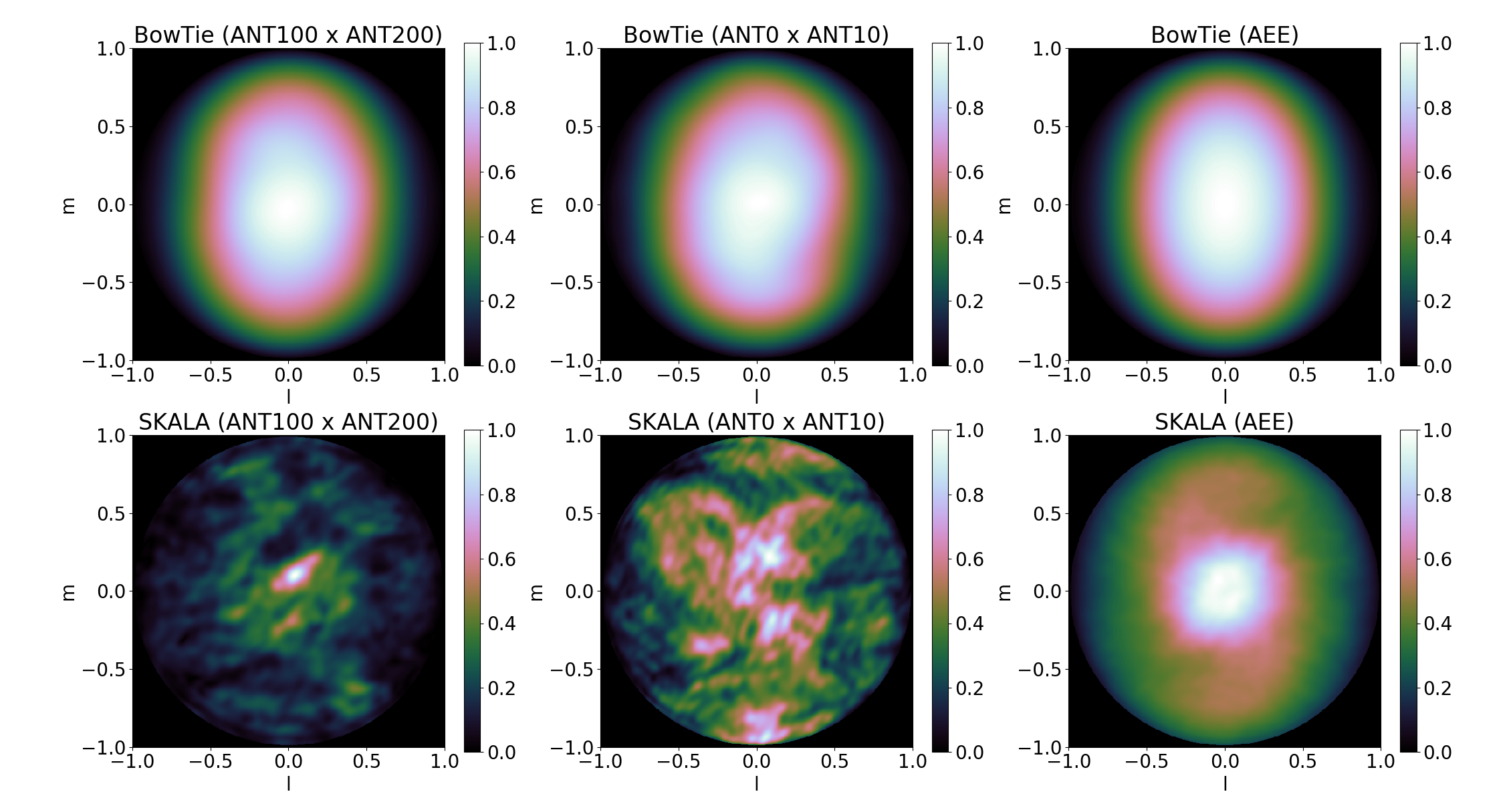}
\caption{Examples of the complex baseline-dependent antenna power pattern formed from randomly chosen antennas (left and centre \added{columns}) vs the average antenna power pattern for the \added{bow-tie} (top \added{right}) and SKALA2 (bottom \added{right}) antennas. Each power pattern is normalised to the maximum power.}
\label{fig:EEPvAEE}
\end{figure*}

In Figure \ref{fig:EEPvAEE} we compare two arbitrarily chosen baseline dependent beam patterns with the AEE power pattern for each antenna design. The AEE pattern is real valued but the baseline dependent beams are complex for which only the magnitude is shown in this figure. As determined in \cite{SKALA_mutual_coupling} we expect to see significant mutual coupling for the SKALA2 antenna\added{,} especially at this frequency of 160MHz. This is exactly what we observe in Figure \ref{fig:EEPvAEE} with the SKALA2\deleted{'s} \added{antennas} exhibiting greater levels of spatial variations compared to the bow-tie antenna. Additionally, the AEE power pattern remains similar to the expected beam shape despite these variations.

\begin{figure}
    \includegraphics[width=\columnwidth]{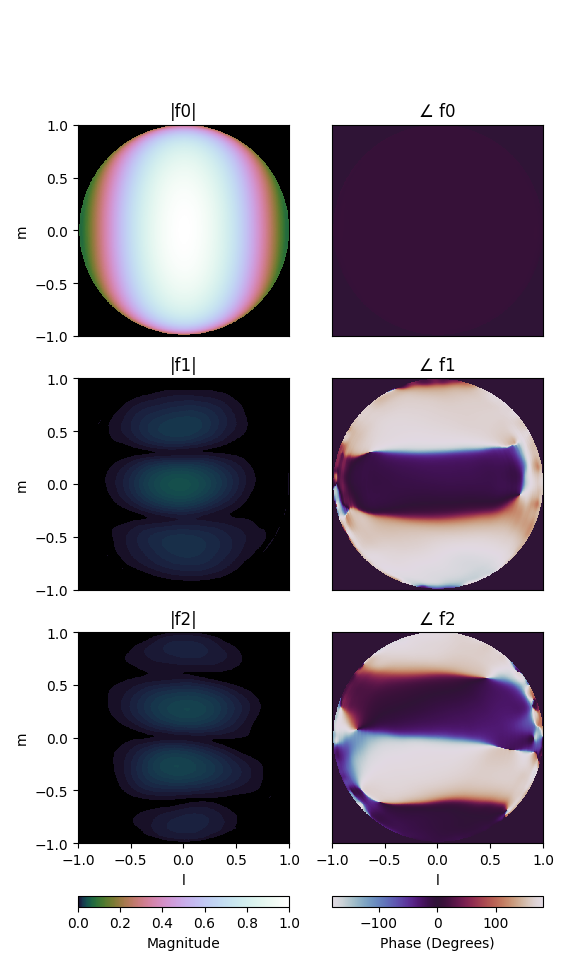}
    \caption{The Magnitude (left) and Phase (right) of the first 3 component functions for the bow-tie antenna.}
    \label{fig:eda_basis}
\end{figure}

\begin{figure}
    \includegraphics[width=\columnwidth]{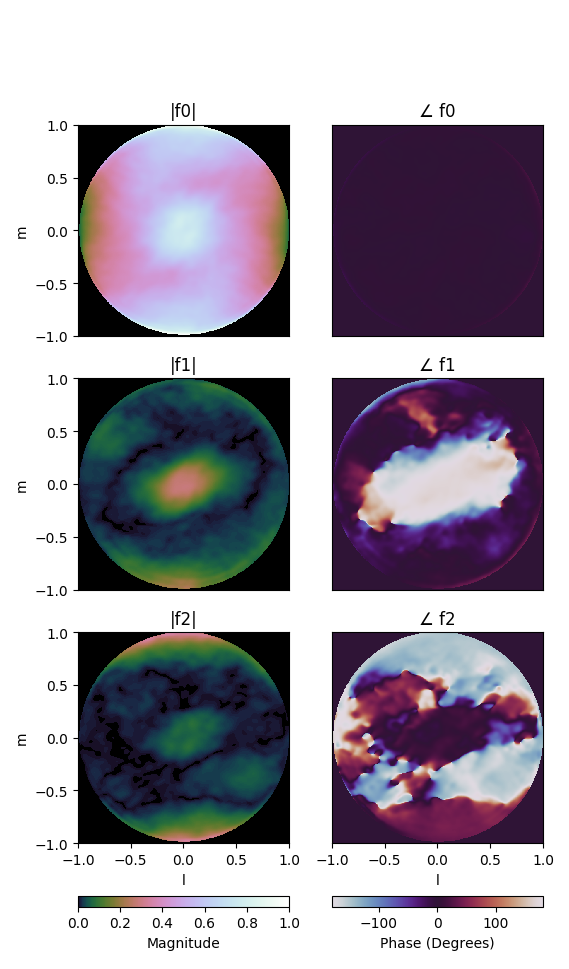}
    \caption{The Magnitude (left) and Phase (right) of the first 3 component functions for the SKALA2 antenna.}
    \label{fig:aavs1_basis}
\end{figure}

Next, in figures \ref{fig:eda_basis} and \ref{fig:aavs1_basis} we show the three largest contributing component functions obtained via the Truncated SVD for the bow-tie and SKALA2 antennas respectively. Since the coefficient vectors are orthonormal, the relative magnitude of the component functions indicate the extent of their overall contribution to the beam model. In the case of the bow-tie antenna\added{,} the main component has a much larger power compared to the 2nd \& 3rd, which is not true for the SKALA2 antenna, this suggests that the higher order components for the bow-tie \added{antenna} contribute less to the final sum. Conversely the higher order components for the SKALA2 \added{antenna} still have a significant contribution\added{. B}ased on the EEP's observed in Figure \ref{fig:EEPvAEE} it is no surprise that the SKALA2 \added{antennas} will likely require more higher order components to accurately represent the model.

\section{Calibration Accuracy}

Modelling the beam patterns using a low-order approximation inherently introduces systematic errors that manifest as a bias within the calibration solutions.
In general\added{,} the quality of the beam model is dependent on the number of components it uses; unfortunately the computational complexity also scales linearly with the number of components\added{,} ultimately leading to a trade-off between calibration accuracy and computational complexity. 
In this section we quantify the calibration accuracy as a function of the number of components used and the incident sky brightness distribution for both the bow-tie and SKALA2 antenna designs. The procedure is outlined as follows:
\begin{enumerate}

\item Generate the sky model: Since the embedded element patterns are simulated at 160MHz we must generate a feasible all sky model for this frequency.
\added{The compact size of an SKA-Low station makes it most sensitive to diffuse large-scale emission that the sky model must properly characterise. Most sky models available at this frequency either do not include diffuse emission, are missing sections of the sky in the southern hemisphere, or are too low in resolution to properly simulate the largest baselines. Instead,} we take the HASLAM all sky map at 408MHz and scale the brightness based on a spectral index of 2.55 \citep{SpectralIndex}.

\added{An} SKA-Low station has a fixed location and each element is always pointed at the zenith\added{.}
\deleted{the range of possible skies are limited.}
\added{Therefore, i}gnoring the \added{S}un and other non-ideal effects such as RFI and the ionosphere, the sky brightness distribution depends only on the Local Sidereal Time (LST). 
For any given LST the HASLAM all sky map is then projected onto image space represented by a $512 \times 512$ grid.
\added{Evaluating the RIME on a discrete grid requires sufficient image resolution to prevent aliasing and therefore prevent any contamination of the visibilities.
The image resolution of $N_p = 512$ was chosen as it is the smallest power of 2 that leads to error in the visibilities no larger than 0.5\% when compared to visibilities computed using a much higher resolution of $N_p = 8192$. 
}

\added{
In this paper we are only concerned about errors introduced as a result of using an approximate beam model and how it interacts with the sky brightness distribution. 
Even if an exact beam model was used, the error in the sky model must be sufficiently small in order to satisfy the SKA-Low requirements for calibration accuracy.
Introducing an approximate beam model will result in an interaction with both the exact sky and the sky modelling error, leading to an additional bias in the calibration solutions. The sky modelling error is unknown but since we have already established that it's comparatively small in magnitude, it can be assumed that it has little influence on the total calibration bias.
Therefore, from this point forward it is assumed that the sky model is free of error as we intend to probe only the systematic errors due to the beam model.
}

\item Generate \added{the noise-free measured coherence matrix ($\mathbf{\tilde{R}}$) using the exact beam}: This matrix represents the ideal measured interferometer response for a noiseless system. It is computed by directly evaluating the RIME in equation \ref{eqn:RIME} using the sky that was previously generated and the exact beam model. 
\added{In practice, the measured coherence matrix is obtained after time averaging but the RIME describes the instantaneous interferometer response. However, it is assumed that the integration time is of the order of 1 second, therefore any time smearing effects should be negligible.}

\item Generate model \added{coherence} matrix ($\mathbf{R}$): This is the matrix under test, it is evaluated using the A-stacking algorithm described by equation \ref{eqn:A-stack} and will therefore contain a bias that will propagate into the calibration solutions. This step is repeated for each beam model under test, i.e. for an increasing number of model components.

\item Compute calibration solution: The calibration solutions are obtained using the formulation described in equation \ref{eqn:Cal} and evaluated through the use of the Lavenberg-Marquart algorithm.

\item Compute calibration bias: The \added{measured coherence} matrix was computed using element gains with unity magnitude and zero phase, any deviation from this expected value represents a systematic bias. 
\added{The magnitude bias is therefore defined by the RMS percentage error between the simulated and expected solutions as follows,}
\begin{align}
    \epsilon_{mag} &= \text{RMS}\Big\{ \frac{||\mathbf{g}||}{1} - 1 \Big\} \times 100 \added{,} \label{eqn:BiasMag}
\end{align}
\added{where $\mathbf{g}$ represents the complex valued calibration solutions for the array.}
\added{The phase bias is defined as the absolute RMS error between the simulated and expected phase solutions,}
\begin{align}
    \epsilon_{phase} &= \text{RMS}\Big\{ \angle ( \mathbf{g} ) - 0  \Big\} \added{.} \label{eqn:BiasPhase} 
\end{align}

\added{Overall, the final data products are the magnitude and phase bias evaluated over a grid of LST's ranging between 0 and 360 degrees, with a step size of 1 degree. For EDA2, the evaluation grid for the number of components range between 0 and 50, with a step size of 1. For AAVS1, the grid ranges between 0 and 250, with a step size of 5 components.}

\end{enumerate}

\subsection{Calibration during the night}

\begin{figure*}
\begin{subfigure}{\columnwidth}
\includegraphics[width=\columnwidth]{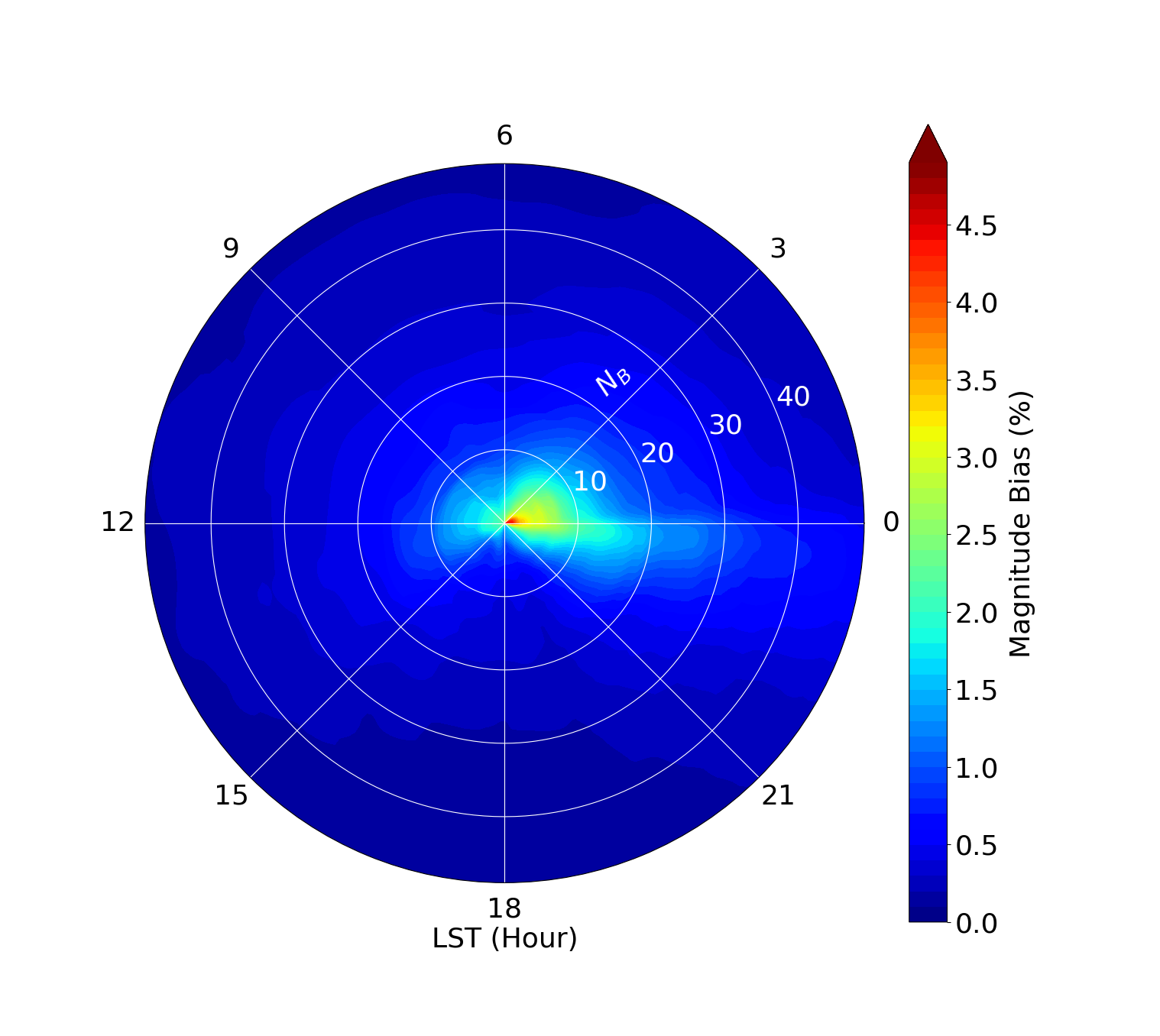}
\caption{ Magnitude bias for bow-tie antenna. }
\end{subfigure}
\begin{subfigure}{\columnwidth}
\includegraphics[width=\columnwidth]{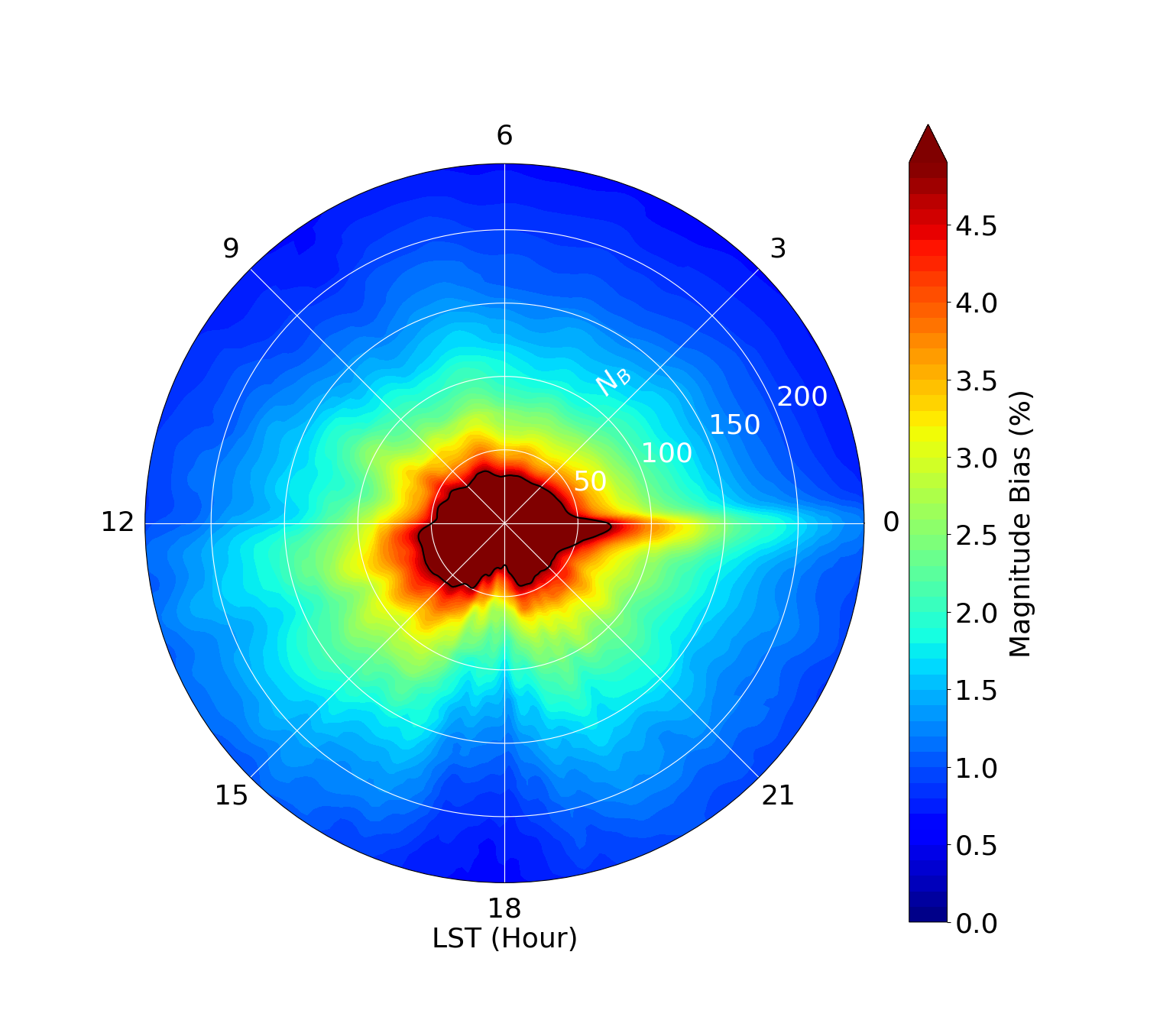}
\caption{ Magnitude bias for SKALA2 antenna. }
\end{subfigure}\par\medskip
\begin{subfigure}{\columnwidth}
\includegraphics[width=\columnwidth]{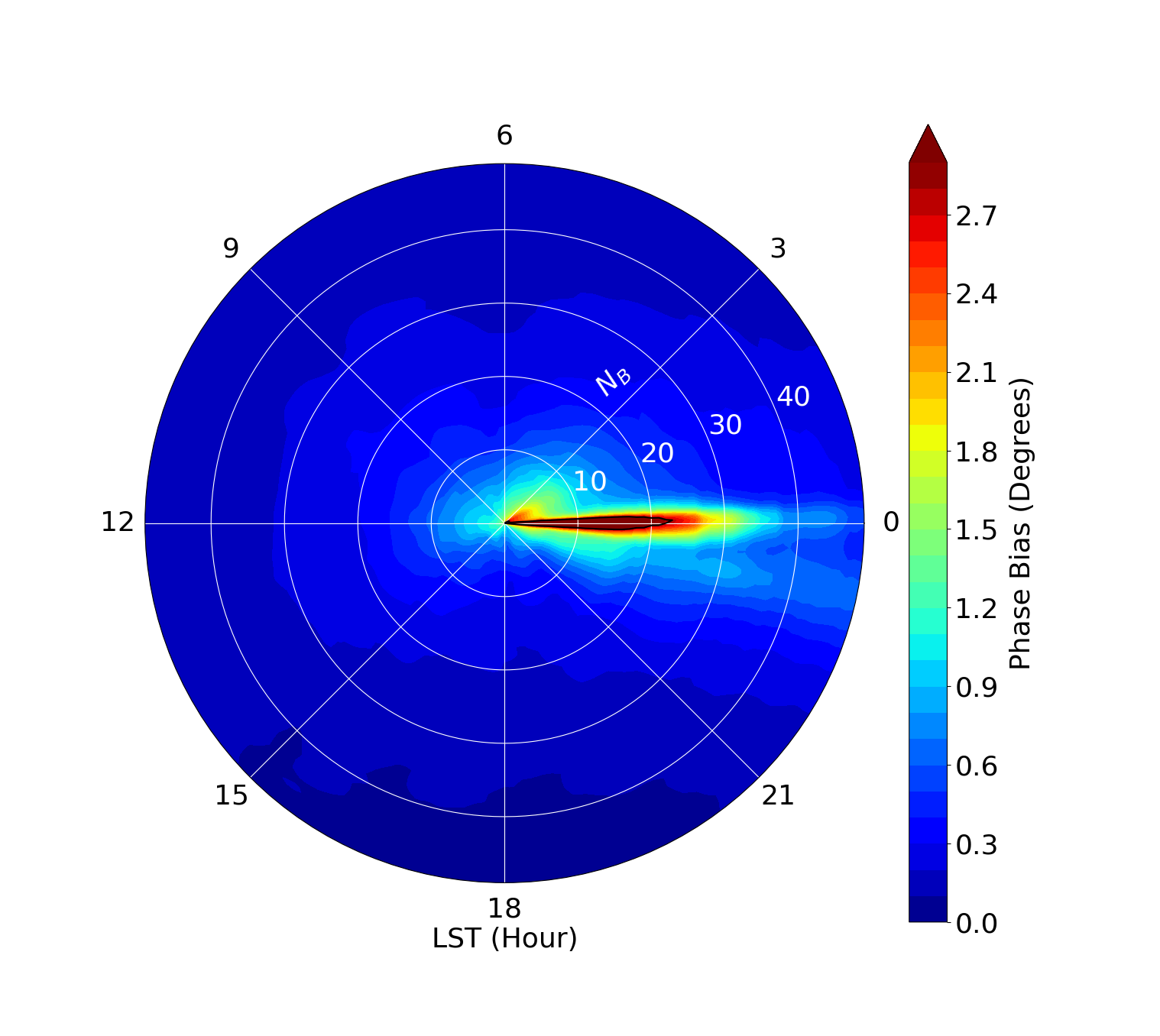}
\caption{ Phase bias for bow-tie antenna. }
\end{subfigure}
\begin{subfigure}{\columnwidth}
\includegraphics[width=\columnwidth]{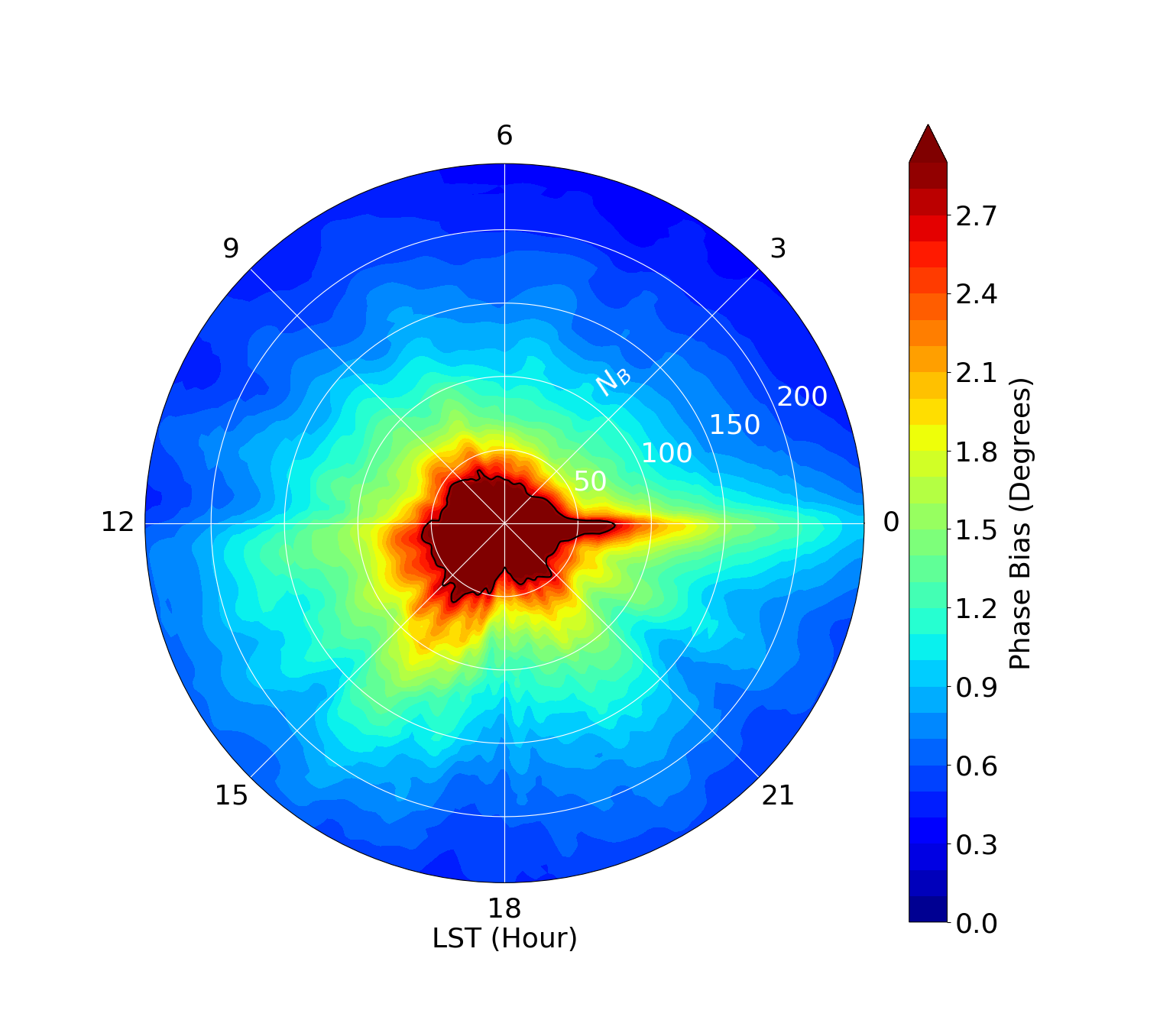}
\caption{ Phase bias for SKALA2 antenna. }
\end{subfigure}
\caption{ The calibration bias for the bow-tie and SKALA2 antenna designs during the night. The angular axis represents the Local Sidereal Time, the radial axis represents the number of model components used, the colour scale indicates the magnitude of the error, and the black contour outlines the bias threshold based on SKA requirements. Note the change in radial-axis scale between the two antenna designs. }
\label{fig:BiasNight}
\end{figure*}

Figure \ref{fig:BiasNight} shows the result of evaluating the calibration bias for all possible sky brightness distributions that could occur at night and for beam models of increasing accuracy. In the figure, the radial axis \added{indicates} the number of components used to represent the beam and the colour-bar indicates the magnitude of \added{the} error. It is \added{clear} that regardless of antenna design\added{,} the \added{LST and therefore the} sky-brightness distribution has a significant effect on the calibration accuracy. It is \added{certainly} the case that there are \deleted{certain} LST's in which the calibration accuracy is particularly poor\added{, namely around 0h LST,} which would prove to be problematic for short calibration intervals.

\deleted{In particular} The models perform generally worse around 0h LST which happens to correspond to the galactic plane setting on the horizon. \added{At this} time there are \deleted{also} very few \deleted{other} strong sources in the main part of the beam \added{apart from the galactic plane}. This effect is most obvious for the bow-tie antenna but it is also evident for the SKALA2 \added{antenna}. Because the SVD minimises the absolute error and not the relative error of the beam model, the relative error is generally worse in directions where the beam is small in magnitude, for this reason the relative modelling error is worse near the horizon. The sky brightness distribution effectively weights the beam model in certain directions more than others\added{.}
Therefore\added{,} when the majority of the sky power is near the horizon it acts to amplify the modelling error in that direction\added{,} which decreases the calibration accuracy. This effect can also be observed between the LST's of 12 and 15 hours for the SKALA2 antenna, this corresponds to the galactic plane rising again which results in most of the sky power \added{being} located at low elevations.

Comparing the two antenna designs it is evident that the SKALA2 antennas are significantly harder to model, requiring more components to achieve similar levels of bias compared to that required by the bow-tie antenna. Note that the radial axis for the SKALA2 plots in Figure \ref{fig:BiasNight} are in fact 5 times larger in scale compared that in the bow-tie plots.

\subsection{Calibration during the day}
A-stacking performs particularly poorly for bright point sources that dominate the sky\added{. T}herefore\added{,} due to the compactness and extreme brightness of the \added{S}un compared to the diffuse background sky, calibration during the day may require special treatment of the \added{S}un. \added{For a smooth and constant valued sky}, the total error after integrating \deleted{over the sky} would effectively average out\added{.} However\added{,} a bright point source will strongly weight the error in one direction, preventing any averaging effect. This effect is made worse if the source is located in a region where the modelling error is \deleted{particularly} poor, such as near the horizon.

Fortunately, as stated in \cite{CAL-DDE}\added{,} FFT based methods such as A-stacking can be combined favourably with the DFT approach\added{,} where the DFT is applied to only the brightest point sources and A-stacking is used only for fainter diffuse objects. The DFT approach is comparatively cheap when applied only to the \added{S}un because it need only be applied to a small number of pixels around it. This hybrid algorithm presents the opportunity to not only circumvent the poor performance of A-stacking during the day but possibly improve the daytime calibration accuracy overall.

Following the same methodology as in the previous section, the calibration accuracy was evaluated over a range of possible skies. However, in this case the \added{S}un was included within the sky model with a fixed declination of 0 degrees and a right ascension of 6h, therefore only LST's 0h through 12h were simulated as the \added{S}un is below the horizon elsewhere.
The flux density of the \added{S}un used within the sky model was determined based on equation 2 from \cite{QuietSun} which approximates the quiet \added{S}un between the frequency range 30 and 350 MHz.
Next, the model \added{coherence} matrix was evaluated using this hybrid method before the calibration solutions and \added{the} resulting bias were computed.

\begin{figure*}
\begin{subfigure}{\columnwidth}
\includegraphics[width=\columnwidth]{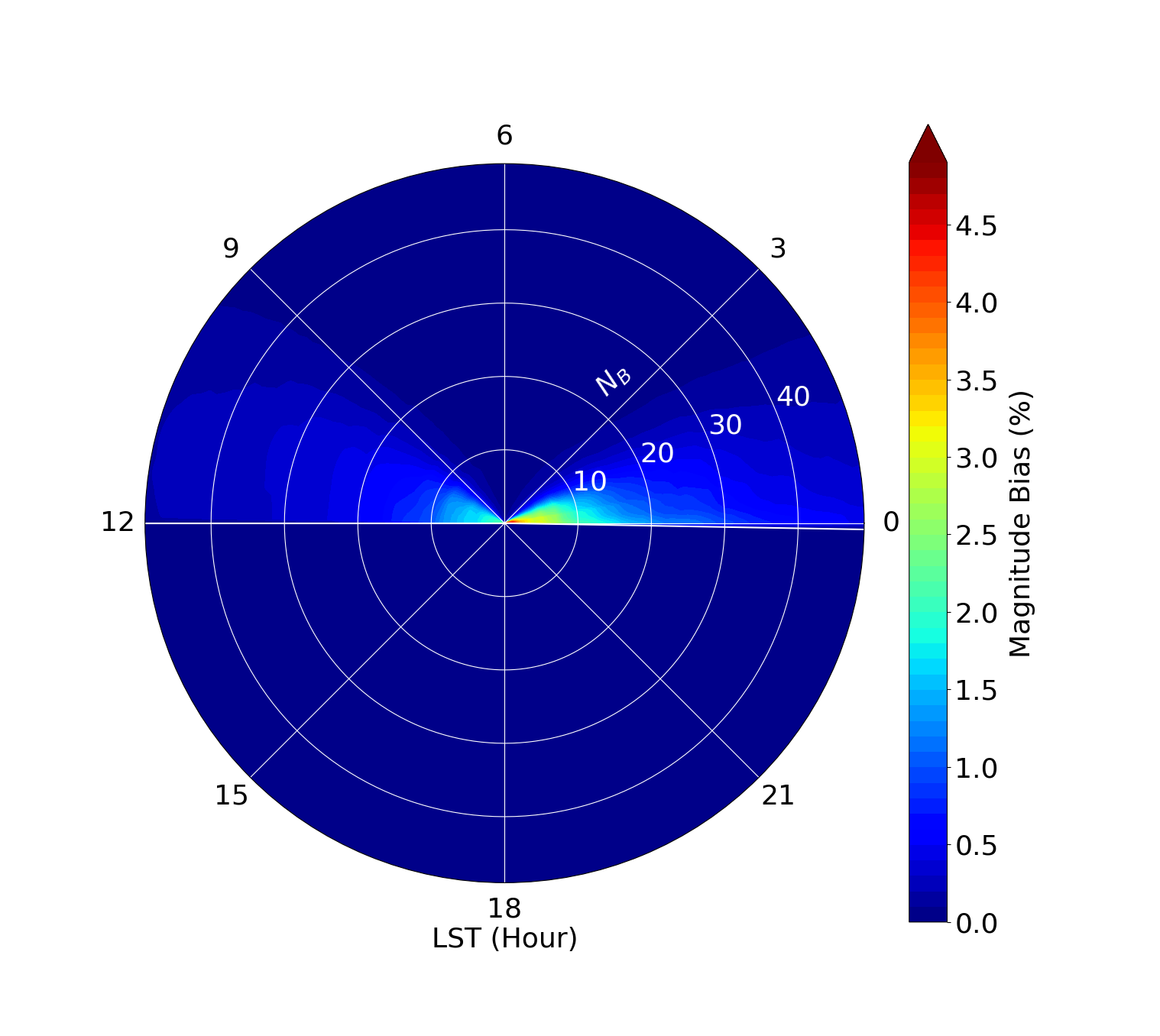}
\caption{ Magnitude bias for bow-tie antenna. }
\end{subfigure}
\begin{subfigure}{\columnwidth}
\includegraphics[width=\columnwidth]{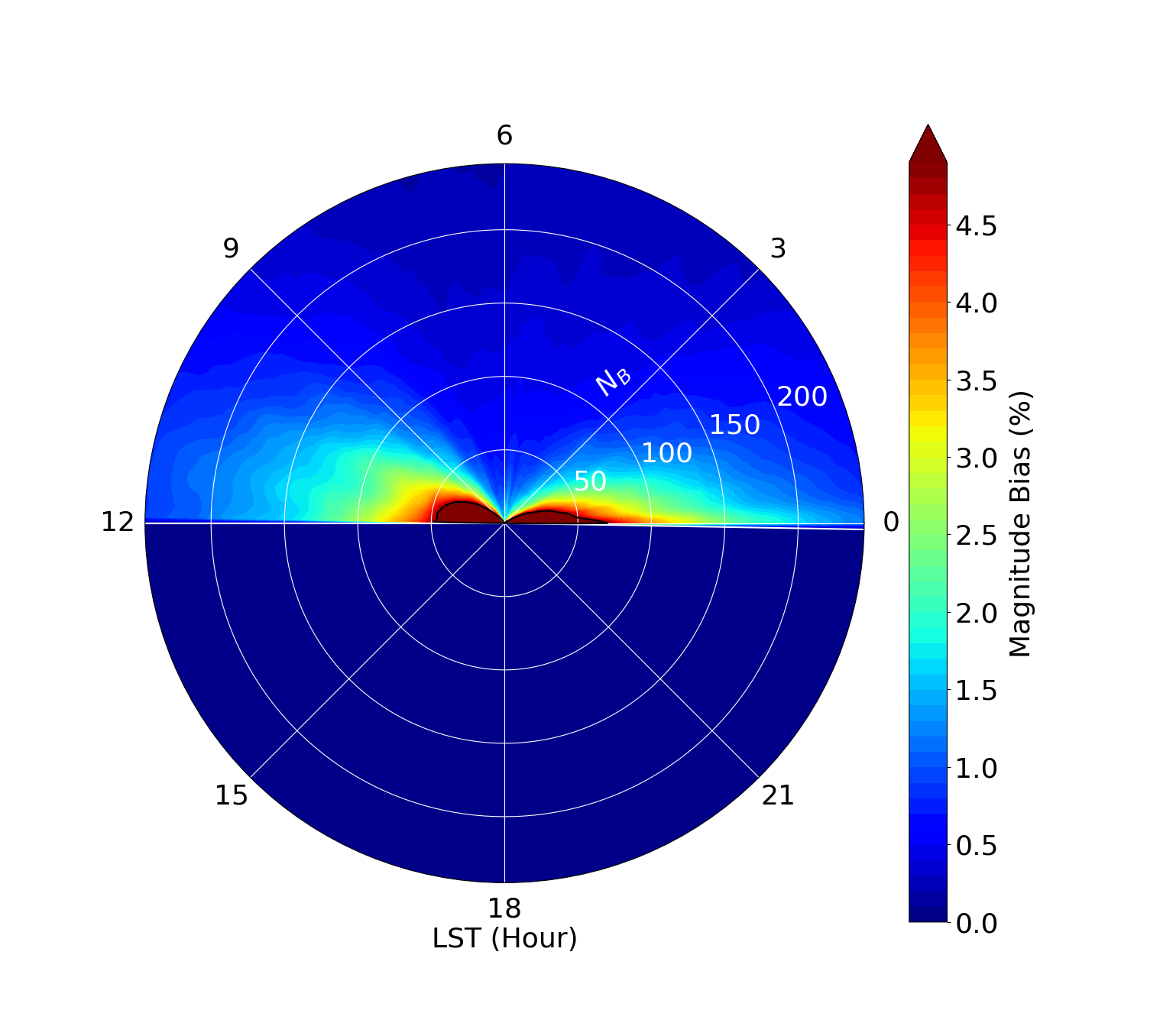}
\caption{ Magnitude bias for SKALA2 antenna. }
\end{subfigure}\par\medskip
\begin{subfigure}{\columnwidth}
\includegraphics[width=\columnwidth]{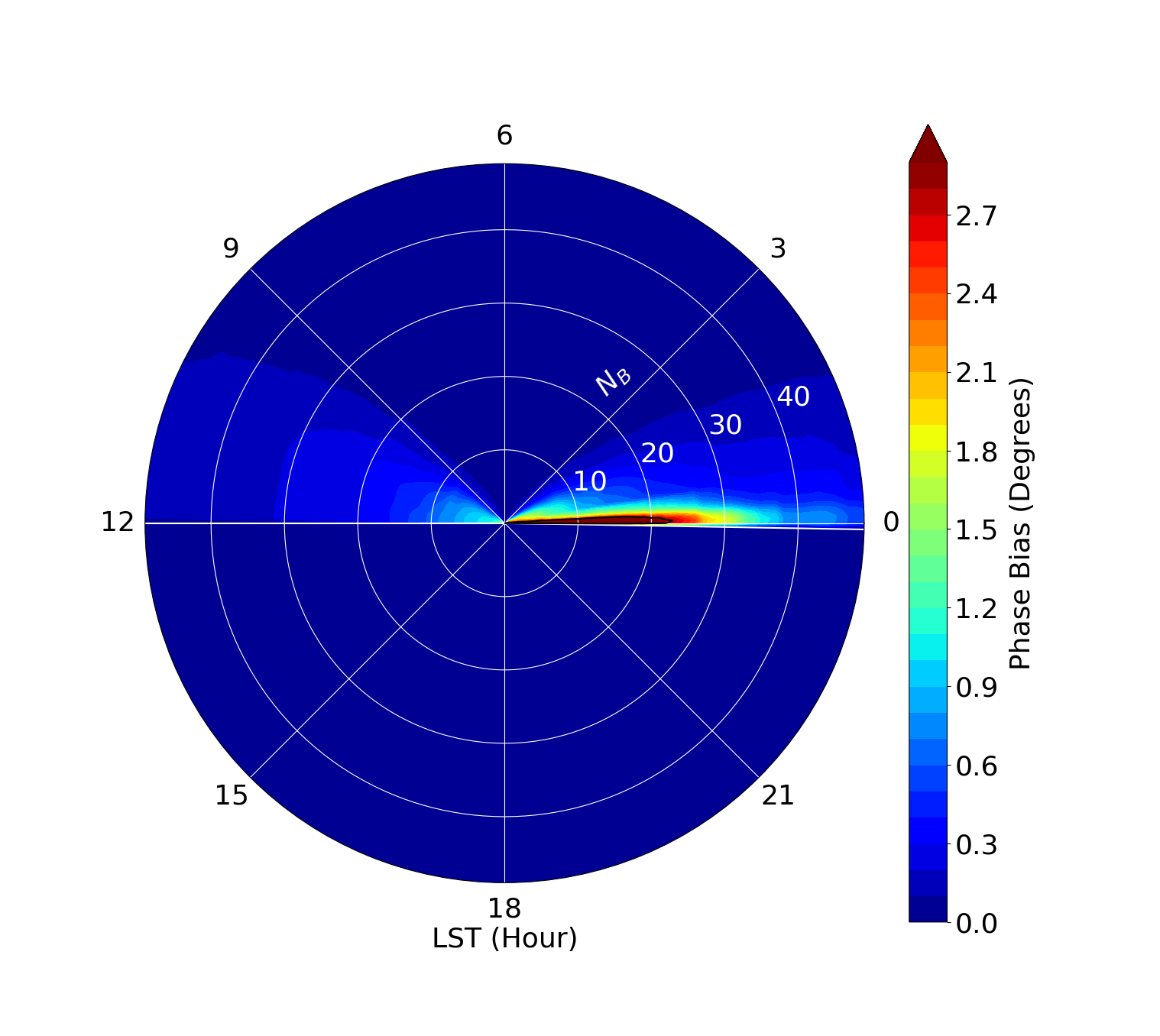}
\caption{ Phase bias for bow-tie antenna. }
\end{subfigure}
\begin{subfigure}{\columnwidth}
\includegraphics[width=\columnwidth]{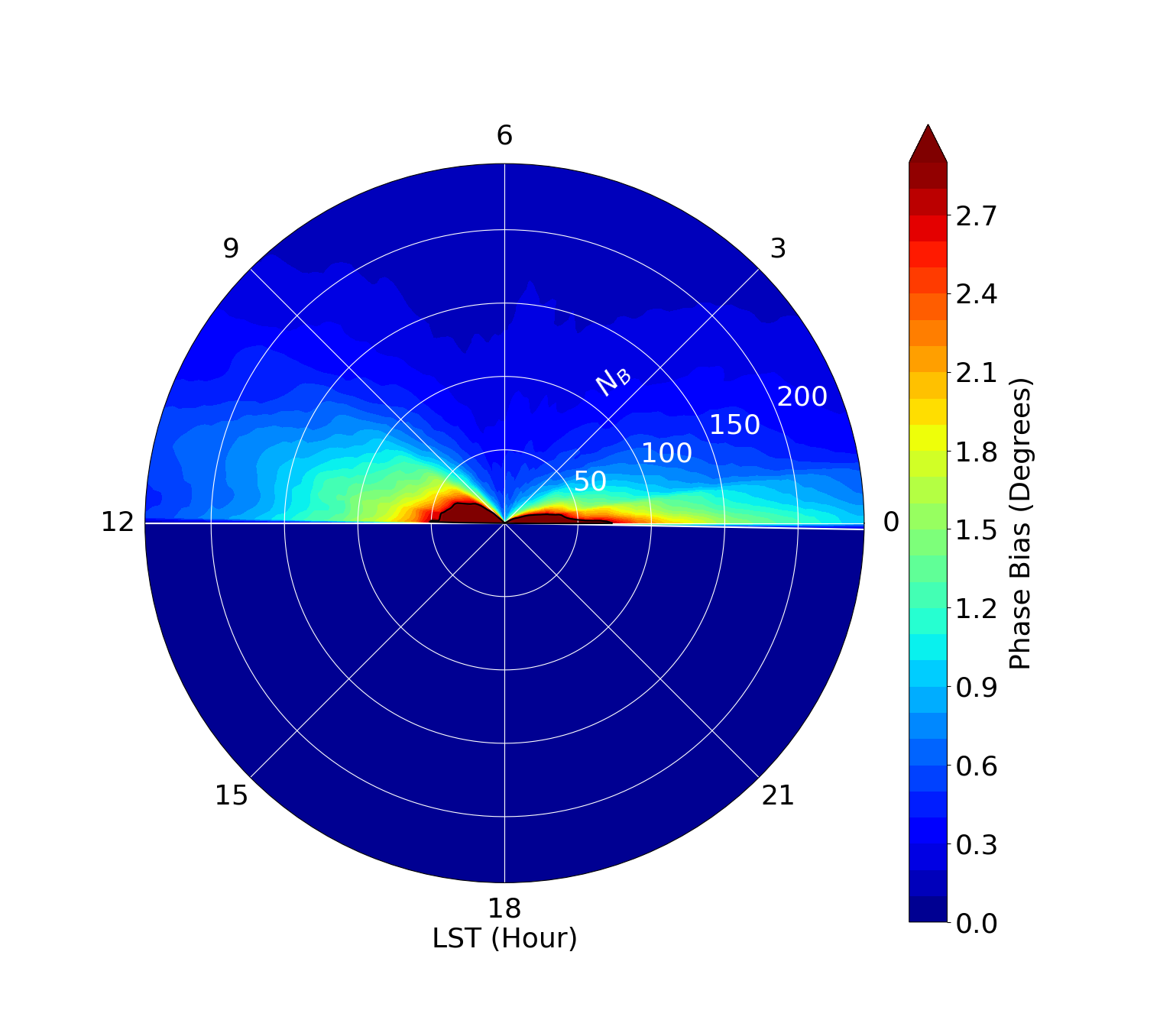}
\caption{ Phase bias for SKALA2 antenna. }
\end{subfigure}
\caption{ The calibration bias for the bow-tie and SKALA2 antenna designs during the day. The \added{S}un is located at RA=90 degrees and DEC=0 degrees. The angular axis represents the LST and hence the sky brightness distribution, the radial axis represents the number of beam model components used and the black contour outlines the bias threshold based on SKA requirements. Note the change in radial-axis scale between the two antenna designs. }
\label{fig:BiasDay}
\end{figure*}

Figure \ref{fig:BiasDay} shows the expected calibration bias evaluated for a typical day over a 12 hour period. The key observation this figure illustrates is that as the \added{S}un passes overhead the calibration bias is significantly reduced, this behaviour is particularly obvious when comparing the SKALA2 antenna between day and night. 
Due to the EEP response of the antennas, which significantly attenuate close to the horizons, the \added{S}un only gradually becomes the most dominant source in the sky as it rises. Therefore at midday, when the \added{S}un is the dominant source the majority of the result is calculated using the DFT method and is therefore error free. At sunrise and sunset the \added{S}un is no longer dominant and therefore the calibration bias is comparable to that seen during the night. Overall, the calibration bias during the day is either equal or less than that observed at the same LST's during the night. The use of a hybrid method when calibrating during the day results in decreased bias overall and a negligible increase in the computational cost.

\section{Meeting the SKA-Low requirements}

The core requirement of the SKA-Low is the ability to achieve thermal noise limited imaging after 1000 hours of integration time\added{. T}herefore\added{,} a strict error budget is defined to limit both systematic and random errors so that this goal can be achieved. The key system requirement of relevance to this paper is SYS\_REQ-2629 named "station beam spatial stability", it states that the RMS percentage difference between the actual and parameterized beam for any beam pointed at zenith angles less than 45 degrees\added{,} after calibration\added{,} should not exceed 0.04\% (at 160 MHz). Derived from this core requirement is requirement LFAA-133 named "relative path stability"\added{; it} limits the maximum amount the receive paths can drift with respect to each other over a 10 minute period to 4.78\% RMS in magnitude and 2.74 degrees RMS in phase (at 160 MHz). While this requirement was designed to limit drift due to random errors over time, it's also implied that systematic sources of error should not exceed this limit either. Overall, the number of components used in a beam model should be chosen such that the bias in the calibration solutions do not exceed the threshold stated above.

One must keep in mind that the systematic error under test in this paper is solely due to the error of modelling the baseline dependent beam patterns, this error is deliberately introduced in an effort to reduce \added{the} computational cost. However, there are several sources of error that must not be neglected, each of which will contribute to the total error budget specified by the SKA-Low requirements. \added{In practice, the maximum limit for this source of error will be even stricter than that quoted previously as the error budget must be shared between all other error sources.}
In general\added{, the systematic} errors arise from three major sources, the accuracy of the embedded element patterns, the accuracy of the sky model and the beam modelling error.

\subsection{Strong sources near the horizon}
\label{sec:horizon}
Fig. \ref{fig:BiasNight}b highlights a general issue, which in this case manifests itself when strong sources are near the horizon.
Simulations predict the largest relative variations in the element beam patterns occur at low elevations. Hence, when a strong source is at low elevation, the received signal is dominated by the variations in the patterns.
This effect proves to be difficult to model accurately, not only are the SVD beam models least accurate at low elevations but \added{the simulated EEP's also have the least certainty near the horizon.}
In addition, strong compact sources at low elevations are more likely to scintillate at low frequencies, which is an additional source of \added{error that arises} from the use of an \emph{a-priori} sky model. Therefore, irrespective of the beam modelling error, the \added{increased uncertainty of other systematic error sources near the horizon} mean that in practice\added{,} the calibration solutions are likely to be least accurate in \added{these} cases.

\noindent\deleted{5.2 Calibration update rate}

Due to \added{the expected} instability of the receive paths within an SKA-Low station it was initially specified that calibration shall be continually evaluated once every 10 minutes. \added{This} may not be feasible due to both the computational cost of regular calibration and the ability to achieve the desired level of accuracy at all LST's. 
As discussed previously, solutions obtained when strong sources are near the horizon should be regarded as unreliable. \added{This} is problematic if calibration is required to have a 10 minute cadence because \added{these} sources can spend significant amounts of time there.
\added{Therefore, it may be necessary to require that} in practice\added{,} the system gains are sufficiently stable over time to allow for less regular calibration.

\subsection{Accuracy requirements}
In figures \ref{fig:BiasNight} and \ref{fig:BiasDay} the receive path stability requirement is denoted by a black contour line, this clearly shows the minimum number of model components that are required as a function of LST. 
The worst case occurs at approximately 0h LST when the galactic plane sets on the horizon, both antennas require the most components at this time and simply finding this maximum point will ensure that the SKA requirement will always be met. Therefore, in the worst case\added{,} the bow-tie antennas require 25 components and the SKALA2 \added{antennas} require 80. \added{Overall, A-Stacking presents a computational advantage over the DFT method, even in the worst case.}

However, as discussed earlier, calibration at LST's in which strong sources are near the horizon should be avoided as \added{it is} likely to lead to increased \added{systematic errors which are very difficult to estimate. Events such as} the rise of the galactic plane can result in an increased calibration bias for up to 3 hours in duration, for example between 12h and 15h LST. 
\added{B}y selecting several favourable LST's roughly 2 to 3 hours apart, the SKALA2 antennas \added{would only require} approximately 50 components \added{to meet the target accuracy limit.}
\added{Similarly, if re-calibration around 0h LST can be avoided,} the bow-tie antenna would only require a single component in order to meet the receive path stability requirement.
\added{Therefore, to reduce the computational complexity even further we require that the minimum re-calibration rate is no more than once every 2 to 3 hours. Which may be necessary irrespective of the beam model used in order to avoid other sources of error from both the EEP's and the sky model at low elevations.}
Since calibration during the day will require no more components than that required at night, day time calibration has no effect on the required number of components. Overall, the number of beam model components required for the bow-tie and SKALA2 antennas is 1 and 50 respectively.

\added{This result is obtained by ignoring other sources of systematic error, this was done in order to probe only the error that arises due to approximating the beam model. In particular, we have assumed that both the simulated EEP's and the sky model are error free, but in practice this won't be the case. The total error budget as specified in the SKA-Low requirements must be shared between all sources of systematic error. Therefore, the actual target accuracy of the beam model is likely to be stricter than what has been considered thus far, ultimately requiring more model components. Therefore, the result of 1 and 50 components for the bow-tie and SKALA2 antennas should be considered as an absolute minimum.
For example, if the maximum allowable error is halved the bow-tie antenna would require approximately 8 components, while the SKALA2 antenna would require approximately 110 components.}

\subsection{Computational Complexity}
The 512 unique stations of the SKA-Low will each require regular calibration for each polarisation\added{,} potentially for every coarse channel in real time. This is a significant computational task \added{which consists} of \deleted{the} capturing approximately 1 second of raw voltage data from the receivers, correlating 256 dual pol antennas and evaluating the model visibilities before computing the final calibration solutions.
The evaluation of the model visibilities can easily be overlooked because typically its complexity is negligible compared to the correlation step\added{,} but this is not necessarily the case for arrays with non-homogeneous element patterns. The complexity of correlating 1 second worth of data for 256 elements of a single coarse channel is \deleted{described by}
\begin{align}
    C_{\text{corr}} &=  \mathcal{O}\Big( \ K B \ \Big) \added{,}
    \label{eqn:C_corr}
\end{align}
where \added{$K$ is the number of baselines and} $B$ is the bandwidth of the channel. Evaluating the RIME using the DFT method can only be considered negligible compared to the correlation step if the following relation holds true\added{,}
\begin{align}
      N_p^2 \ll B \added{.}
\end{align}
Since the image size ($N_p \times N_p$) must be at least $512 \times 512$ pixels in order to achieve sufficient \deleted{integration} accuracy, the computational complexity of the \added{DFT} method falls within an order of magnitude compared \added{to} the task of correlating a single course channel with a bandwidth of 781.25kHz. Overall, the computational cost of evaluating the RIME can be significant and shouldn't be overlooked. 

Methods such as the AEE method would result in a major reduction in computational cost, however\added{,} in the case of the SKALA2 antenna it cannot satisfy the accuracy requirements. Fortunately, the A-stacking method provides a middle ground where a trade-off can be made between accuracy and complexity. The A-stacking method can be considered negligible compared to the correlation step if
\begin{align}
      N_B N_{gW}^2 \ll B \added{,}
\end{align}
which holds true since the de-gridding\added{/W-projection} kernel $N_{gW}$ is small \added{(approximately 20)} and the number of components \added{ required ($N_B$) ranges from 1 to 110}. Overall, the A-stacking method can simultaneously meet the accuracy requirements while avoiding the \added{substantial} computational burden of directly evaluating the RIME.

\section{Conclusions} 

The calibration of interferometer arrays with non-homogeneous element patterns can be \added{substantially} more complex compared to calibrating a typical homogeneous array.
In the context of the SKA-Low, station level calibration proves to be a significant challenge due to the extent of mutual coupling exhibited between elements\added{,} resulting in non-homogeneous patterns. The application of the A-stacking method within this paper enabled a compromise to be made between the accuracy and computational complexity of array calibration. The strict requirements surrounding the accuracy of the SKA-Low enforces a minimum limit on the computational cost required to achieve it. It was shown that this limit depends on the non-homogeneity of the element patterns in addition to the sky brightness distribution and the calibration update rate. 

The minimum required number of model components was evaluated for the two antenna designs under consideration for the SKA-Low. It was shown that the SKALA2 antenna required \added{at least} 50 components compared to the single component required for the bow-tie antenna. This difference is explained by the extent of non-homogeneity that each antenna design exhibits.
Since the computational complexity scales linearly with number of components, it follows that the computational cost for the SKALA2 array is approximately 50 times the cost required by a bow-tie array. While this result was arrived at using a number of assumptions and for a single polarisation and frequency, it can be reasonably concluded that the SKALA2 antenna has at least an order of magnitude increased cost associated with it.
Considering the total number of unique SKA-Low stations and the number of frequency points calibration will need to be evaluated for, this increased computational cost could prove to be quite significant.

Furthermore, the computational cost was shown to increase in situations where there are strong sources near the horizon, in particular the rise and fall of the galactic plane which can last for up to 3 hours at a time. 
\added{A further decrease in computational cost can be achieved if calibration can be avoided in these situations.}
However, it remains to be shown \added{whether} the receive path variability of an SKA\added{-Low} station is stable enough over time to \added{achieve this}.

Overall, it has been shown through simulations\added{,} that the A-stacking method can be employed favourably to meet a target accuracy while minimising the computational cost of calibrating an interferometer array with non-homogeneous element patterns. In future work, this method should be demonstrated using a real system to determine its viability in practice.

\section{Data Availability}
The data underlying this article will be shared on reasonable request to the corresponding author.



\bibliographystyle{mnras}
\bibliography{refs} 







\bsp	
\label{lastpage}
\end{document}